\documentclass[12pt]{article}
\usepackage{graphicx}
\begin{document}
\LaTeX
\bigskip
\begin{center}
The reflection spectra of the Bechgaard salts: a theoretical determination 
\end{center}

\medskip
\begin{center}
V. \v Celebonovi\'c
\medskip

Institute of Physics,Pregrevica 118,11080 Zemun-Beograd

Serbia
\end{center}
\medskip 
\begin{center}
e-mail:vladan@ipb.ac.rs  
\end{center} 
\date{\today} 



\begin{abstract}
The aim of this contribution is to analyze to some extent the reflectivity of family of organic salts known as the Bechgaard salts. The reflectivity will be calculated using a well known theoretical framework . Since this calculation demands as input data the electrical conductivity and the susceptibility,existing results for these quantities were used. The main motivation for this calculation was to attempt to "tie up" the reflectivity with various parameters of the Bechgaard salts in the Hubbard model. As these materials are quasi one dimensional,it is hoped to extend the results of this contribution to systems of higher dimensionality. 
\medbreak
Details of the calculation leading to an expression for the reflectivity will be discussed. It is non linear,and shows the dependence of the reflectivity on various material paremeters (within the Hubbard model), such as the transfer integral or the band filling. The position of the region where the reflectivity tends to zero will also be estimated as a function of the model parameters. Finally,some possibilities of extension of the results obtained to 2D systems are briefly mentioned.  
\medbreak  
\end{abstract}
PACS {78.67.-n}
 \footnotetext[1]{Published in Acta Phys.Polonica {\bf A112},949 (2007).} 
\section{Introduction}
The family of the organic conductors with the generic chemical formula 
$(TMTSF)_{2}X$ was synthetized in 1980.,and later named the Bechgaard 
salts [1]. In this formula, the symbol
$(TMTSF)_{2}$ denotes a complex called di-methyl-tetra-selena-fulvalene,
and $X$ is one of various kinds of anions,which can be added to the complex.
A few examples of the anions are $X=FSO_{3},ClO_{4},NO_{3}..$. Already early work,reviewed in [2], has shown that the electrical conductivity of the Bechgaard salts can not be described within the standard theory of metals. A new theoretical framework was obviously needed,and the choice "by default" was the Hubbard model. The calculation of the electrical conductivity of these materials using the so called "`memory function"' method has recently been reviewed in [3]. The essential conclusion of [3] is that the calculation discussed there is in semi-quantitative agreement with experimental data. In this paper,the reflectivity of the Bechgaard salts will be calculated,using well known equations from optics [4],and results for the electrical conductivity of these materials [3].   

\section{The method of calculation}
The propagation of an electromagnetic wave in a nonmagnetic material is determined by the dielectric function $\epsilon(\omega)$ and the refractive index $N(\omega)$; $N(\omega)=[\epsilon(\omega)]^{1/2}$ [4]. Both of these quantities can be represented as complex functions: 
\begin{equation}
	\epsilon(\omega)= \epsilon_{R}(\omega)+i\epsilon_{I}(\omega) 
\end{equation}
and
\begin{equation}
	N(\omega)=n(\omega)+iK(\omega) 
\end{equation}
 The function $K(\omega)$ denotes the extinction coefficient. 
It can be shown [4] that
\begin{equation}
	\epsilon_{R}(\omega)= n^{2}-K^{2} 
\end{equation} 

\begin{equation}
	\epsilon_{I}(\omega)= 2nK 
\end{equation} 
The dielectric function $\epsilon(\omega)$ is related to the susceptibility $\chi(\omega)$ by: $\epsilon=1+4\pi\chi$. The susceptibility of the Bachgaard salts is a complex function and has recently been determined in [3]. 

The reflectivity $R(\omega)$ is defined as the following ratio

\begin{equation}
R(\omega)=\frac{(n-1)^2+K^{2}}{(n+1)^2+K^2}	
\end{equation}
  
\section{The results}
In order to introduce material parameters into eq.(2.5) one has to insert in it the electrical conductivity;this can be achieved through the relation
\begin{equation}
\epsilon_{R}(\omega)+i\epsilon_{I}(\omega)	=1+i\frac{4\pi(\sigma_{R}(\omega)+i\sigma_{I}(\omega)) }{\omega}
\end{equation}
where the conductivity $\sigma$ has been expressed as a complex function: $\sigma(\omega)=\sigma_{R}(\omega)+i\sigma_{I}(\omega)$. Similarily,$\epsilon(\omega)=\epsilon_{R}(\omega)+i\epsilon_{I}(\omega)$. Combining the expressions for the real and immaginary parts of $\epsilon$ and $\sigma$,some simple algebra leads to the following:
\begin{equation}
n=\frac{2\pi\sigma_{R}}{\omega K} 
\end{equation} 
and 
\begin{equation}
K^{4}+K^{2} (1-\frac{4\pi\sigma_{I}}{\omega})- n^{2}=0
\end{equation}
Inserting eq.(3.2) into (3.3) and solving for $K$ gives
\begin{equation}
K^{2}_{1,2} = \frac{4\pi\omega\sigma_{I}-\omega^{2}\pm[(4\pi\omega\sigma_{R})^{2}+(\omega^{2}-4\pi\omega\sigma_{I})^{2}]^{1/2}}{2\omega^{2}}
\end{equation}
Inserting results for $n$ and $K$ into eq.(2.5) leads to the final expression for the reflectivity
\begin{equation}
R=\frac{4\pi^{2}\sigma_{R}^{2}+K^{2}(K^{2}+\omega^{2}(1-2n))}{4\pi^2\sigma_{R}^{2}+K^{2}(K^{2}+\omega^{2}(1+2n))}	
\end{equation}

The real and immaginary parts of the conductivity are related to the corresponding parts of the susceptibility by
\begin{equation}
	\sigma_{I}=\frac{\omega_{P}^{2}}{4\pi \omega}\left(1-\frac{\chi_{R}}{\chi_{0}}\right)
\end{equation} 
and 
\begin{equation}
	\sigma_{R}=\frac{\omega_{P}^{2}\chi_{I}}{4\pi \omega \chi_{0}}
\end{equation}


It has been shown ([3] and later work)  that the real part of the susceptibility has the form of a sum: 
\begin{equation}
	\chi_{R} = \sum_{i} \frac{A_{i}}{\omega+q_{i} t} 
\end{equation} 
while the immaginary part can be expressed as: 
\begin{equation}
	\chi_{I}=\sum_{i} \frac{A_{i}}{\pi}\frac{\omega}{(q_{i} t)^{2}}\frac{1}{1-(\omega/q_{i} t)^{2}} \ln(\frac{\omega}{q_{i} t})^{2} 
\end{equation}

In this expression,$A_{i}$ contains various material parameters,as defined in the Hubbard model,and $q_{i}$ is a numerical constant. The expressions for the real and immaginary parts of the susceptibility contain the chemical potential of the electron gas,discussed in [3]. Using the fact that $(1+x)/(1+y)\cong1+x-y-xy+y^{2}...$ ,and using eq.(3.2) one could transform eq.(3.5) into a practically more useful expression for $R$.It can be shown that the reflectivity is approximately given by

\begin{equation}
	R\cong1-\frac{2\omega}{K\pi\sigma_{R}}+\frac{1}{2}\frac{K\omega}{(\pi\sigma_{R})^{3}}
\end{equation}

 
\section{Discussion}
Expression (3.5) is the final result for the reflectivity of the Bechgaard salts.It is obviously nonlinear. A considerably simplified form is given by eq.(3.10). Inserting into it eqs.(3.4) and (3.6)-(3.9) into (3.5) or (3.10) gives the dependence of the reflectivity of the Bechgaard salts on their basic parameters (as defined within the Hubbard model): the Hubbard $U$,the hopping energy $t$,the band filling $n$ (implicitely,through the chemical potential),the frequency $\omega$,the plasma frequency $\omega_{P}$ and of course the inverse temperature $\beta$.  

A detailed analysis of eq.(3.5) and the influence of various material parameters on the behaviour of $R$ would be too long for presentation here. However,eq.(3.10) makes some interesting conclusions relatively easy to obtain. 

A frequently occuring question when analyzing the reflectivity of any optical material is whether the reflectivity can under some conditions become equal to zero,or become arbitrarily small. 

Using eqs.(3.2) and (3.6) it follows that $R\cong0$ for 
\begin{equation}
\omega=\frac{2K(\pi\sigma_{R})^3}{(2\pi\sigma_{R})^2-K^{2}}
\end{equation}
Inserting the appropriate expressions for the conductivities and $K$ one could determine the range of material parameters for which $R\cong0$. 

Experiments on Bechgaard salts are usually performed under low temperatures. As an illustration of the behaviour of the reflectivity of these materials calculated in the present paper,the sums in eqs.(3.8) and (3.9) were limited to 9 terms,and then inserted into eq.(3.10). The value of the reflectivity was calculated for the same values of parameters as in [3]: $N=150$;$s=1$;$U=4 t$;$\omega_{P}=12t$;$\omega=3t$;$\chi_{0}=1/3$. The value of the reflectivity was normalized to the value at the point ( $T=100 K$,$t=0.001 eV$). The temperature dependence of the reflectivity thus calulated for the band filling of $n=0.8$ is shown in the following figure. 


\begin{figure}
\includegraphics [width=11.5cm]{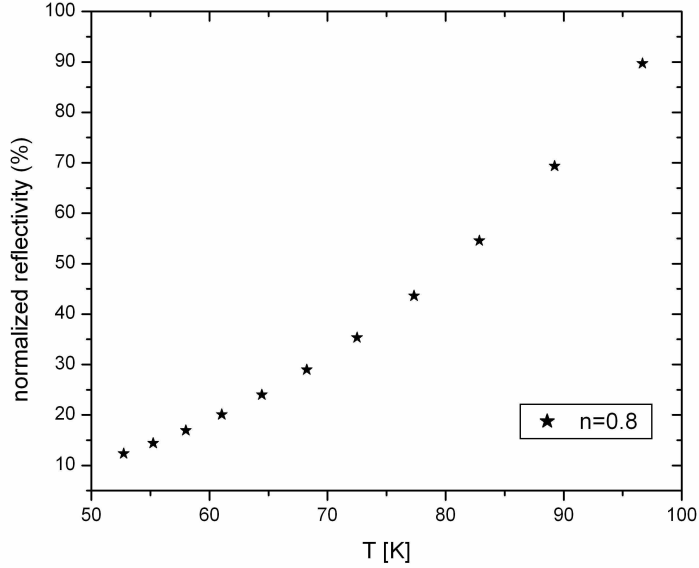} \caption{Normalized reflectivity for n=0.8} \label{Fig1}
\end{figure} 

It is clear from the figure that for the chosen set of parameters,and within the approximation used in the present paper, the reflectivity decreases with the decrease of the temperature. Whether or not this can be stated as a general conclusion,will be the subject of future work.     



 
\medbreak
\section{Acknowledgement}
This paper was prepared as a part of the research project 141007 financed by the Ministry of Science of Serbia.

\medbreak


\end{document}